\def\etal{et al}
\RequirePackage{fix-cm}
\documentclass[aps,prb,showpacs,superscriptaddress,a4paper]{revtex4-1}
\usepackage{amsmath}
\usepackage{float}
\usepackage{subfigure}
\usepackage{color}
\usepackage{color, colortbl}
\definecolor{Gray}{gray}{0.75}
\definecolor{LightCyan}{rgb}{0.50,1,1}
\usepackage[section]{placeins}
\usepackage{amsfonts}
\usepackage{amssymb}
\usepackage{graphicx}
\begin {document}
\author{Banasree Sadhukhan}\email{banasree.rs@presiuniv.ac.in}
\author{Arabinda Nayak}\email{arabinda.physics@presiuniv.ac.in}
\affiliation{Department of Physics, Presidency University, 86/1College Street, Kolkata 700073, India}
\author{Abhijit Mookerjee}\email{abhijit@bose.res.in}
\affiliation{Professor Emeritus, S.N. Bose National Center for Basic Sciences, JD III, Salt Lake, Kolkata 700098, India}

\title{\bf Effect of doping on the electronic properties of Graphene and T-graphene : A theoretical approach}
\date{\today}
 
\begin{abstract}
In this communication we present together four distinct techniques for the study of electronic structure of solids : the tight-binding linear muffin-tin orbitals (TB-LMTO), the real space and augmented space recursions and the modified exchange-correlation. Using this we investigate the effect of random vacancies on the electronic properties of the carbon hexagonal allotrope, graphene, and the non-hexagonal allotrope, planar T graphene. We have  inserted random vacancies at different concentrations, to simulate disorder in pristine graphene and planar T graphene sheets. The resulting disorder, both on-site (diagonal disorder) as well as in the hopping integrals (off-diagonal disorder), introduces sharp peaks in the vicinity of the Dirac point built up from localized states for both hexagonal and non-hexagonal structures. These peaks become  resonances with increasing vacancy concentration.  We find that in presence of vacancies, graphene-like linear dispersion appears in planar T graphene and the cross points form a loop in the first Brillouin zone similar to buckled T graphene  that originates from $\pi$ and $\pi$* bands without regular hexagonal symmetry.  We also calculate the single-particle  relaxation time, $\tau(\vec{q})$ of $\vec{q}$ labeled quantum electronic states which originates from  scattering due to presence of vacancies, causing quantum level broadening. .
\end{abstract}

\date{\today}

\maketitle

%%%%%%%%%%%%%%%%%%%%%%%%%%%%%%%%%%%%%%%%%%%%%%%%%%%%%%%%%%%%%%%%%%%%%%%%%%%%%%%%%%%%%%%%%%%%%%%%%%%%%%%%%%%%%%%%%%%%%%%%%%%%%%%%%%%%%

\section{Introduction}

                            Graphenes are allotropes of carbon forming almost planar, one atom thick sheets. The carbon atoms in graphene are arranged in a two-dimensional (2D) \cite{2d} honeycomb lattice, and are the building blocks of many multi-functional  nano-materials. Ever since its discovery there has been keen interest in graphene not only because of its unconventional low-energy behavior : half-integer Hall effect, ultra-high electron mobility, low resistivity, ballistic electronic conductivity and massless carriers \cite{massless}, but also because of its potential technological applications \cite{apl1} - \cite{apl5}.  The unique  properties of graphene arise from its linear dispersion  around the Dirac point. This  is attributed to the crossing of  $\pi$ and $\pi^*$ bands at the $K$ and $K'$ points of the Fermi energy level in the reciprocal space. The opposite cones at the K point in the band structure arise from the graphene's hexagonal crystal symmetry. In trying to justify its Dirac-like massless excitations many authors have used  the relativistic Dirac equation to study graphene. However, given that carbon has atomic number
 six, the average speed of excitations near the Fermi level is far less than the speed of light and direct relativistic effects are  unlikely. The phenomenon of linear dispersion can arise from the  geometry of  its ion-core arrangement. There can be no significant relativistic corrections in carbon. The Dirac point and linear dispersion can be obtained from a  standard Schr\"odinger equation using a transparent real space approach. It has been conjectured that this behavior is a result of the real space hexagonal symmetry of the ion-core lattice. However, even this conjecture is doubtful. Enyashin and Ivanovskii \cite{hexa} constructed 12 artificial 2D carbon networks varying hybridizations among the carbon atoms. He also examined their stability and electronic structures. They found that  Dirac fermions can exist in strained graphene without a regular hexagonal symmetry \cite{hexa} - \cite{hexa2}. T graphene  with its tetra and octa rings is an example of this \cite{hexa}. This allotrope composed of a periodic array of tetragonal and octagonal (4,8) carbon rings can be produced by cleaving two adjacent atomic layers in a body-centred cubic C$_8$ or trigonal C$_4$ along the (001) direction. T graphene is energetically metastable and dynamically stable \cite{hexa2}. 
                            
                            We should note that a minor local modification in graphene can lead to the formation of a T graphene nucleus. This is shown in Fig.\ref{TG}. If we remove two neighboring carbon atoms and re-bond the dangling bonds so produced, then a minor relaxation leads to a topological defect which forms  the seed of T graphene. Such  topological ``impurities" stretch over more than sixteen sites.  Liu $\etal$ \cite{hexa2} have proposed two new stable forms of a 2D tetra-symmetrical carbon allotrope: planar T graphene and buckled T graphene. They have reported that buckled T graphene has Dirac-like fermions due to its two types of non-equivalent bonds. The crossing of $\pi$ and $\pi$* bands forms a loop with tetra symmetry in the Brillouin zone.  Planar T graphene is metallic though it has $\pi$ and $\pi$* bands near the Fermi level, which in agreement with the previous reports \cite{hexa2}, \cite{kim} - \cite{chun}. Crossing $\pi$ and $\pi$* bands are not sufficient conditions to produce linear dispersion relation near the Fermi surface. Planar T graphene has one sublattice in a unit cell whereas, buckled T graphene has two, like graphene. These characteristics in buckled T graphene are believed to be another important factor for inducing linear dispersion near Fermi surface. Buckled and planar T graphenes   have comparable formation energies but the latter is thermodynamically more  stable  below around T=900 K \cite{hexa2}. From a fully relaxed calculation  and  simple bonding arguments, Kim $\etal$ \cite{kim} have shown that   buckled T graphene cannot be distinguished from the planar one. This indicates that buckled T graphene is not stable even at high temparatures and it gains stability under spontaneous structural transformation into the planar T graphene. Buckled T graphene has similar structure to that of planar T graphene except two adjacent square lattices are tilted, one in the
upward z-direction and the other in the downwards z-direction. The 
z-range is about 0.55 \AA .   Several phenomena such as T graphene-substrate interaction, absorption of molecules and doping, changes the planar T graphene's behavior from metallic to a Dirac dispersion character at room temperatures. C.S Liu $\etal$ \cite{chun} have demonstrated that Lithium decorated T graphene becomes a feasible nanosensor and exhibits  high sensitivity to carbon monoxide without hexagonal symmetry like pristine graphene.
            
     The aim of this communication will be to use a combination of the tight-binding linear muffin-tin orbitals method and the real space  and the augmented space recursion methods to try to answer the questions raised. The TB-LMTO augmented space recursion (ASR) allows us to study the effects of geometry, chemistry {\it and} disorder. This will be the focus in this paper. 
                            
 \subsection{Dealing with disorder}
 
                                   Disorder is ubiquitous in graphene and graphene like materials. The existence of these defects suggest the possibility of magnetic moments and their ordering in graphene. These localized electronic states at zigzag edges and vacancies leads to an extreme enhancement of the spin polarizability \cite{pala} - \cite{oleg2}. The model calculations suggest that the magnetic moments will form in the vicinity of these defects. Disorder causes the  energy bands  to gain width related to the disorder induced quasi-particle relaxation time $ \tau(\vec{q})$  \cite{am2} - \cite{config}. Theoretically,  randomness (disorder) is introduced through an external parameter. In a super-cell approach, the complex part in energy usually not calculated but given ``by hand" just like a external parameter to introduce disorder. An alternative  approach  is through  mean-field theories, the most successful among which is the coherent potential approximation (CPA). But CPA is a single site formalism and any kind of multi-site correlated  randomness or extended defects cannot be addressed by it. This brings us to the third approach. This problem can be dealt with, within the hierarchy of the generalizations of the CPA such as nonlocal CPA \cite{nlcpa}, the special quasirandom structures (SQSs) \cite{sqs}, the locally self consistent multiple scattering approach (LSMS) \cite{lsms}, and few other successful approaches such as the traveling cluster approximation (TCA) \cite{tca}, the itinerant coherent potential approximation (ICPA) \cite{icpa} and the tight binding linear muffin-tin orbitals (TB-LMTO) based augmented space recursion technique (ASR) \cite{am2} - \cite{config}. Over the years, the ASR has proved to be one of the most powerful techniques, which can accurately take into account the effects of correlated fluctuations arising out of the disorder in the local environment. The CPA, the non-local CPA, the itinerant CPA (ICPA) and the traveling cluster approximation (TCA) are all special cases derivable from the augmented space formalism (ASF). The formalism also allows us to account for various types of scattering mechanisms in a parameter-free way. 

                                 In this work, we address the effect of random vacancies \cite{peres} - \cite{inv} on the electronic properties of graphene and planar T graphene. The calculation of average density of states (ADOS) in both graphene and T graphene with binary alloy disorder (vacancies) by the recursion method reveals the presence of resonant peaks  at the Dirac point \cite{inv}. The shapes and positions  of the resonant peaks  essentially depend on the impurity concentration and the on-site potential. We have also carried out a study on the spectral properties and energy  band dispersion of both hexagonal and non hexagonal carbon allotropes with and without vacancies. We have calculated disorder induced single-particle relaxation times ($ \tau(\vec{q}) $)  from the self-energy \cite{Doniach}, \cite{Doniach1} due to scattering from vacancies.
 
 %****************************************************************************************

Disorder poses another problem. For example, to study a disordered binary alloy, the experimentalist works on   
one macroscopic sample and not $2^N$ possible random configurations.
The equality of  spatial   and configuration averaging under specific conditions is known as the 
theorem of spatial ergodicity \cite{config}. The augmented space formalism
assumes at the outset that spatial ergodicity holds for the disordered system under consideration. We shall also assume that spatial ergodicity holds in the
systems that are under study. 

As an example, we shall begin with the simplest model : the atoms in our
solid sit on a geometric perfect lattice, but the lattice points are occupied by two types of atoms A and B. This bulk alloy has concentrations of A and B as $x$ and $1-x$. The Hamiltonian is :

\begin{eqnarray}
\textbf{H} & = & \sum_{\vec{R}L}\left[ \varepsilon^A_L m_{\vec{R}} + \varepsilon^B_L (1-m_{\vec{R}})\right]   \textbf{P}_{\vec{R},L}+  
 \sum_{\vec{R}L}\sum_{\vec{R}^\prime\ne \vec{R},L^\prime}
\left[\Delta_L^A m_{\vec{R}} + \Delta_L^B (1-m_{\vec{R}})\right]  S_{\vec{R}L,\vec{R}^\prime L^\prime }\ldots\nonumber\\
& & \left[\Delta^A_L m_{\vec{R}^\prime}
 + \Delta^B_L (1-m_{\vec{R}^\prime})\right] \textbf{T}_{\vec{R}L,\vec{R}^\prime L^\prime} \nonumber\\
 & & \phantom{X}\nonumber \\
 &\ =\ & \sum_{\vec{R}L}\left[ \varepsilon^B_L + \delta\varepsilon_L\  m_{\vec{R}} \right]   \textbf{P}_{\vec{R},L}+ 
 \sum_{\vec{R}L}\sum_{\vec{R}^\prime\ne \vec{R},L^\prime}
\left[\Delta_L^B  + \delta\Delta_L\ m_{\vec{R}})\right]  S_{\vec{R}L,\vec{R}^\prime L^\prime }\ldots\nonumber\\
& & \left[\Delta^B_L 
 + \delta\Delta_L\ m_{\vec{R}^\prime})\right] \textbf{T}_{\vec{R}L,\vec{R}^\prime L^\prime} \nonumber\\
 & & \phantom{X}\nonumber\\
 &\ =\ & \textbf{H}^B + \sum_{\vec{R}L} \delta\varepsilon_L\  m_{\vec{R}}   \textbf{P}_{\vec{R},L}+ \sum_{\vec{R}L}\sum_{\vec{R}^\prime\ne \vec{R},L^\prime}
\left[  \Delta^B_L S_{\vec{R}L,\vec{R}^\prime L^\prime } 
  \delta\Delta_L\ m_{\vec{R}^\prime}+\right.\ldots\nonumber\\
& & \left.\ldots\qquad\delta\Delta_L\ m_{\vec{R}} S_{\vec{R}L,\vec{R}^\prime L^\prime } \Delta^B_L\ + 
 \delta\Delta_L\  S_{\vec{R}L,\vec{R}^\prime L^\prime }
 \delta\Delta_L\ m_{\vec{R}} m_{\vec{R}^\prime} \right] 
 \textbf{T}_{\vec{R}L,\vec{R}^\prime L^\prime}\nonumber\\
 & & \phantom{X}\nonumber\\
 & \ =\ &\textbf{H}^B + \delta\textbf{H}[\{m_{\vec{R}}\}] 
\end{eqnarray}
The $m_{\vec{R}}$ randomly take the values 1 or 0 according to whether
the site $\vec{R}$ is occupied by A or B type of atom with probability
$x$ or $y\ =\ 1-x$.

The augmented space formalism now adopts ideas from the theory of measurements.
With each measurement ${\cal M}_k$ we associate an operator 
${\bf M^k}\in \Phi^k$ such that the actual measured results are the
eigenvalues of ${\bf M^k}$ : $\{m^k_\mu\}$ and its spectral density is  their probability density. The process of measurement then can be 
described as projection of an unknown configuration state to an 
eigenstate of ${\bf M^k}$. These eigenstates span the `configuration space' $\Phi^k$.  Using the recursion method in the space $\Phi^k$ we
generate a basis $\{\vert\theta^k_n\rangle\}$ in which the representation of ${\bf M^k}$ is a Jacobian matrix. 

\begin{eqnarray*}
\vert\theta^k_1\rangle &&\mbox{the choice of the initial state is open}\\
\vert\theta^k_2\rangle &=& {\bf M}^k \vert\theta^k_1\rangle - \alpha_1
\vert\theta^k_1\rangle \\
\vert\theta^k_n\rangle &=& {\bf M}^k \vert\theta^k_{n-1}\rangle -\alpha_{n-1}\vert\theta^k_{n-1}\rangle - \beta^2_{n-1}\vert\theta^k_{n-2}\rangle
\end{eqnarray*}
with
\begin{equation} \alpha_n = \langle \theta^k_n \vert {\bf M}^k\vert\theta^k_{n}\rangle
\qquad \beta_n = \langle \theta^k_n \vert {\bf M}^k\vert\theta^k_{n-1}\rangle\end{equation}

The augmented space theorem\cite{am2}$^-$\cite{am1} tells us that the configuration average of any functional of
a set $\{ m^k\}$\ (k=1,2$\ldots$ ) is 
 \begin{equation} \ll {\bf F}(\{m^k\})\gg  = \langle\Theta_1\vert \widetilde{\bf F}( {\widetilde{\bf M}}^k\} ) \vert\Theta_1 \rangle
 \end{equation}
 
 where, if {\bf F}$\in\ \Psi$, then $\widetilde{\bf F}( \{\bf \widetilde{M}^k\} )
 \in \Psi\otimes\prod^\otimes \Phi^k$ and $\vert\Theta_1\rangle \ =\ 
 \prod^\otimes \vert \theta^k_1\rangle$. The augmented operator $\widetilde{F}$ is constructed by simply replacing the random variable $m^k$ by the operator $\widetilde{\bf M}^k$ and taking the matrix element in
 the `average' state $\vert\Theta_1\rangle$. 

For a set of independent binary random variables $\{m^k\}$ taking values 0 and 1 with probablities $x$ and $y=1-x$, corresponding operator {\bf M}$^k$ is of rank 2 and has a Jacobi representation :

\begin{equation} \textbf{M}^k \in \Phi^k = \left(\begin{array}{cc}
                                 x & \sqrt{xy} \\
                                 \sqrt{xy} & y \end{array}\right)  
\end{equation}                                 
  The rank of the `configuration space' $\Phi = \prod^\otimes \Phi^k$ is
  $2^N$ as it should.
  $ \widetilde{\bf M}^k = {\bf I}\otimes{\bf I}\ldots\otimes {\bf M}^k \otimes\ldots $  and the `average state' $\vert\Theta_1\rangle =
  \prod^\otimes \vert\theta^k_1\rangle$                                

The augmented Hamiltonian is :

\begin{eqnarray}
 \widetilde{\bf H} & = & \sum_{L} \langle\epsilon_L\rangle \sum_{\vec{R}}
 \hat{\bf I}^{\vec{R}}\otimes{\bf P}_{\vec{R}L} + \sum_L \delta\epsilon_L\ \sum_{\vec{R}}\widehat{\bf M}^{\vec{R}}\otimes{\bf P}_{\vec{R}L} 
+\ldots\nonumber\\
& &\ldots \sum_{\vec{R}L} \sum_{\vec{R}^\prime\ne \vec{R},L^\prime} \left[\langle\Delta_L\rangle S_{\vec{R}L,\vec{R}^\prime L^\prime} \langle\Delta_L\rangle \right] \tilde{\bf I} \otimes {\bf T}_{\vec{R}L,\vec{R}^\prime L^\prime} + \ldots\nonumber \\
& & \ldots 
  \sum_{\vec{R}L} \sum_{\vec{R}^\prime\ne \vec{R},L^\prime} \left[\delta\Delta_L S_{\vec{R}L,\vec{R}^\prime L^\prime} \langle\Delta_{L'}\rangle \right]\hat{\bf I}^{\vec{R}^\prime}\otimes 
\widehat{\bf M}^{\vec{R}}  \otimes {\bf T}_{\vec{R}L,\vec{R}^\prime L^\prime} +\nonumber\\
 & & 
\ldots  \sum_{\vec{R}L} \sum_{\vec{R}^\prime\ne \vec{R},L^\prime} \left[
  \langle\Delta_L\rangle  S_{\vec{R}L,\vec{R}^\prime L^\prime}
\delta\Delta_{L'} 
  \right]\widehat{\bf M}^{\vec{R}^\prime}\otimes\hat{\bf I}^{\vec{R}}  \otimes {\bf T}_{\vec{R}L,\vec{R}^\prime L^\prime} +\nonumber\\
& & \ldots  \sum_{\vec{R}L} \sum_{\vec{R}^\prime\ne \vec{R},L^\prime} \left[\delta\Delta_L  S_{\vec{R}L,\vec{R}^\prime L^\prime} \delta\Delta_{L^\prime}\right]\widehat{\bf M}^{\vec{R}}\otimes \widehat{\bf M}^{\vec{R}^\prime} \otimes {\bf T}_{\vec{R}L,\vec{R}^\prime L^\prime}
\end{eqnarray}

The configuration averaged Green function is given by :

\begin{equation}  \ll G_{\vec{R}L,\vec{R}L}(z)\gg = \langle \vec{R}L \otimes \Theta_1\vert (z\tilde{\bf I}-\widetilde{\bf H})^{-1} \vert \vec{R}L
\otimes \Theta_1\rangle 
\end{equation}

This can be obtained through the recursion method starting from a state :
$ \vert \vec{R}L \otimes\Theta_1\rangle$. The relevant approximation is 
the terminator (after N steps) which is obtained from the behavior of $\{\alpha_n,\beta_n\}$ $n\leq N$.

  The augmented space technique allows us to go beyond the local mean field theories like the coherent potential approximation and describes the extended disorder like short-ranged clustering or ordering as well as long-ranged disorder like randomly placed Stone-Wales defect chains   forming in a graphene background as shown Fig.\ref{SW}. Formation of such long-ranged random defects have been observed by Monte-Carlo simulations in our group \cite{sw1} - \cite{sw2} and will be reported in detail in a subsequent communication. Moreover, if the atomic sizes of the alloy constituents are very different, chemical disorder can also lead to local random lattice distortion. In an earlier work, we had shown how the ASR can deal with systems with local structural disorder in non-isochoric CuBe alloys \cite{cube}. The aim of this work is not only a qualitative description of how disorder affects electronic properties, but to be able to predict quantitatively what this effect leads to.

               Both graphene and T graphene are  two-dimensional networks of carbon atoms, with two and four non-equivalent carbon per unit cell respectively. This is clearly depicted in Fig.\ref{unitcell}. The space group of graphene is P6/mmm with lattice constant 2.46 ${\rm \AA}$  whereas, planar T graphene has space group p4mm with lattice constant 3.42 ${\rm \AA}$. There are many different ways of producing disordered graphene including both intrinsic as well as extrinsic. Intrinsic ways may include surface ripples and topological defects. Extrinsic disorder comes in the form of vacancies, adatoms, quenched substitutional atoms, and extended defects, such as edges and cracks, produced on irradiation of graphene. Such disorder can  considerably modify system's properties \cite{exp1} - \cite{exp5}. In this context available experimental data support the theoretical views . Here we create vacancies by randomly removing carbon atoms from the lattice.

The augmented space formalism (ASF) was introduced by Mookerjee \cite{am1} to deal configuration averages for disordered systems. It has been well known that the effect of disorder fluctuations is dominant at lower dimensions where mean-field approaches are inaccurate. When we are dealing with planar disordered  graphene-like systems we have to perform the configuration averaging in an efficient way. Augmented space recursion technique (ASR) provides a powerful and  accurate tool for studying disorder averaging \cite{config}.

In this work we have used augmented space formalism (ASF) \cite{am2} - \cite{am5} combined with the recursion technique of Haydock $\etal$ \cite{recur} - \cite{recur1}. The recursion method \cite{recur1} is a purely real space   approach for the study of systems where
Bl\"och periodicity fails. It is an ideal method, through the local density of states, to study extended defects like a random nucleation of a 4-8 graphene 
ring in a pristene graphene background.  The augmented space method accurately describes effects of local disorder beyond mean field theories.

 Mathematically, a new, countable, orthonormal basis set $\vert n \gg$ is generated in which
the augmented Hamiltonian is tridiagonal and is constructed  through a three term recurrence formula :

\begin{eqnarray*}
\vert 1\gg &  = & \sqrt{\frac{1}{N}}  \sum_{\vec{R}} \exp\{i\vec{q}\cdot\vec{R}\}  \ \vert\vec{R}\otimes\{\emptyset\}\gg\nonumber \\ 
\vert n+1\gg & = & \widetilde{\bf H}\vert n\gg - \alpha_{n}\vert n\gg - \beta^2_{n-1} \vert n-1\gg 
\end{eqnarray*}

\[
\alpha_n = \frac{\displaystyle \ll n  \vert \widetilde{\mathbf H}\vert n
\gg}{\displaystyle \ll n\vert n\gg} \quad,\quad     \beta^2_n = \frac{\displaystyle \ll n\vert n\gg}{\ll n-1\vert n-1\gg	} 
\]

This leads to :

\begin{equation}
\ll G(\vec{q},z)\gg = \frac{\displaystyle 1}{\displaystyle z-\alpha_1-\frac{\displaystyle \beta_2^2}{\displaystyle z - \alpha_2-\frac{\displaystyle \beta_3^2}{\quad\quad         \frac{\displaystyle \ddots}{\displaystyle  z-\alpha_N-\beta_{N+1}^2 T(z)}}}}
\end{equation}

Notice that while in a single configuration, there is no lattice translation symmetry in the Hamiltonian, if disorder is uniform, then
the full augmented space does. We can therefore talk about a configuration averaged spectral function.
 
\begin{eqnarray}
A({\vec{q}},z) & = & -\frac{1}{\pi}\ \Im m \ll G({\vec{q}},z) \gg \nonumber \\
\end{eqnarray}

Here $\ll...\gg$ indicates a configurational averaged quantity in case of a disordered systems. 

 The problem in any numerical
calculation is that we can deal with only a finite number of operations. In the
recursion algorithm, we can go up to a finite number of steps and if we stop the recursion at
that point, it would lead to exactly what we wish to avoid in the super-cell technique. The analysis of the asymptotic part of the continued fraction is therefore of prime interest to us. This is the
``termination" procedure discussed by Haydock and Nex \cite{haydock}, Luchini and Nex \cite{nex}
, Beer and Pettifor \cite{pettifor} and in considerable detail by Viswanath and M\"uller \cite{vs1} - \cite{vs4}. This terminator $T(z)$ which accurately describes the far environment, must maintain the herglotz analytical properties. We have to incorporate not only the singularities
at the band edges, but also those lying on the compact spectrum of $H$. The nature of the
terminators is deduced from the way in which the recursion coefficients $\{\alpha_n,\beta_n\}$
behave as $n\rightarrow\infty$ and the known singularities of the spectral density. The asymptotic behavior is deduced from the information obtained from the finite calculations and no artificial periodicity or external smoothenings are required. The parameters of the terminator are estimated from the asymptotic part of the continued fraction coefficients calculated from our recursion. We had carried out thirty recursion steps for both graphene and T graphene. For the asymptotic part we have used the  Viswanath-M\"uller termination appropriate for infra-red divergences. In this way both the near and the far environments are accurately taken into account. The real part of the self-energy is an energy shift due to the disorder, and the imaginary part provides the life-time effects.

\begin{figure}[h!]
\centering

\includegraphics[width=8cm,height=5cm]{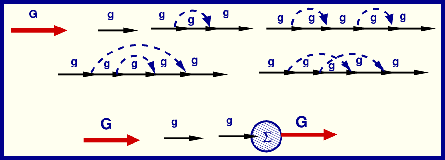}
\caption{(Color online) Dyson Equation because of disorder scattering. $\bf G $ is the single particle Green function, $ \bf \Sigma$ is the electron self energy of the coupled electron-impurity system  .\label{Dys}}
\end{figure}

 Feynman scattering diagrammatic approach we can obtain a Dyson equation as shown in Fig.\ref{Dys}.

\begin{eqnarray} \ll G(z)\gg & = & g(z) + g(z) \Sigma(z) \ll G(z) \gg \nonumber\\
                 \Sigma(z)   & = &  g^{-1}(z)- \ll G(z)\gg^{-1} \ =\ \Sigma_R(z) - i \Sigma_I(z)
                 \end{eqnarray}
                 
Here $g(z)$ is the Green function for the system without disorder $\Delta = 0$. $ \Sigma_I(\vec{q},E) $ is the disorder induced smearing  \cite{Doniach} of a $ {\vec{q}} $ labelled quantum state with energy E($\vec{q}$). These lifetimes have been experimentally measured both in photonic and Raman spectroscopy where a similar phenomenon is observed in phonons.

                                  We choose the symmetric $\Gamma$-K (Armchair) direction i.e. $C<10>$ rather than $\Gamma$-M (Zigzag) direction i.e.  $C<11>$ of graphene and  $\Gamma-X$ and $\Gamma-M$ direction of planar T graphene \cite{hexa2}, \cite{bct1} - \cite{bct2} for carrying out recursion. For a random representation of carbon with vacancies the Hamiltonian was derived self-consistently from the DFT-self-consistent tight-binding linear muffin-tin orbitals augmented space recursion (TB-LMTO-ASR) package developed by our group \cite{ASR} - \cite{sg1}. In the next step, N$^{th}$ Order Muffin Tin Orbital (NMTO) method was used to construct the low-energy Hamiltonian. The Hamiltonian is consisted with only ‘active’ C-$p_z$ states of both graphene. The NMTO method carries out massive downfolding starting from the all-orbital local density approximation (LDA) calculation. This energy selection procedure enables a down-folded Hamiltonian defined in the basis of a minimal set of active orbitals. The off-diagonal elements t(R) are shown in the Table \ref{tab}. Please note that the overlap decays with distance. In this work we have truncated at the nearest neighbour distances. The results agree quite well with the same calculations using other techniques : like the pseudo-potentials (PP) or linear augmented space (LAPW). As our calculational technique is not common, this check with standard results in limiting cases (here, without disorder) was essential. We have used the Cambridge Recursion Library to accurately locate the band edges and hence constructed the
necessary terminator for each angular momentum projected continued fraction. All the calculations are done at T=0K.

\begin{table}[h!]
\centering
\begin{tabular}{|c|cccc|}\hline
(eV) & $\varepsilon$ & t$(\chi_1)$ & t$(\chi_2)$ & t$(\chi_3)$  \\ \hline 
& -0.291   & -2.544  & 0.1668   & -0.1586\\
\hline
\end{tabular}
\caption{Tight-binding parameters of Hamiltonian generated by NMTO. $\chi_n$ refer to the n-th
nearest neighbour vector on the lattice. \label{tab}}
\end{table}

\section{Results and discussion}

\subsection{Electronic structure of ordered graphene and planar T graphene }

\begin{figure}[h!] 
\centering
\includegraphics[width=4.6cm,height=4.7cm,angle=270]{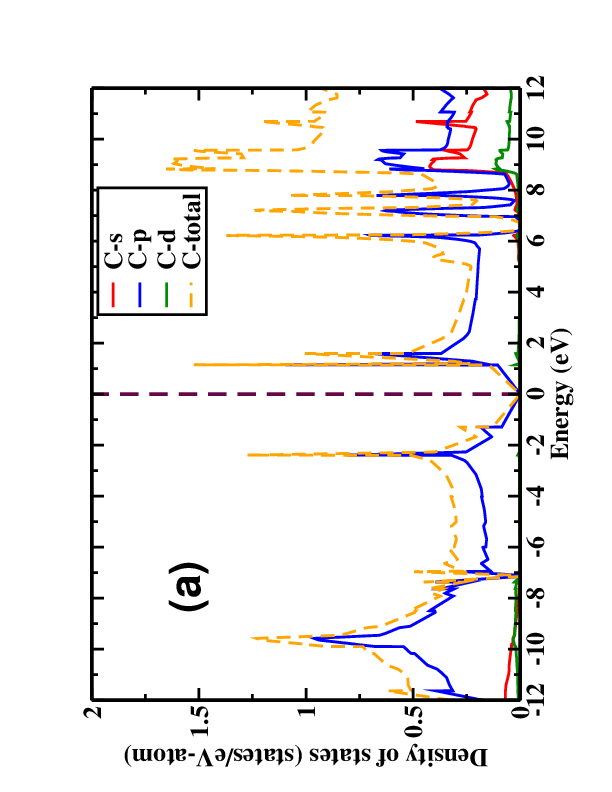}
\includegraphics[width=4.6cm,height=4.7cm,angle=270]{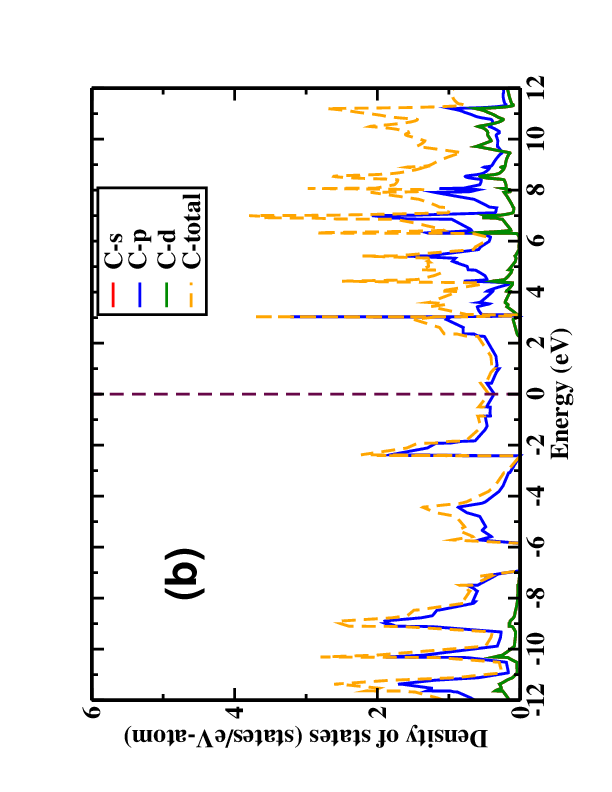}\\
\includegraphics[width=3.5cm,height=4.5cm,angle=270]{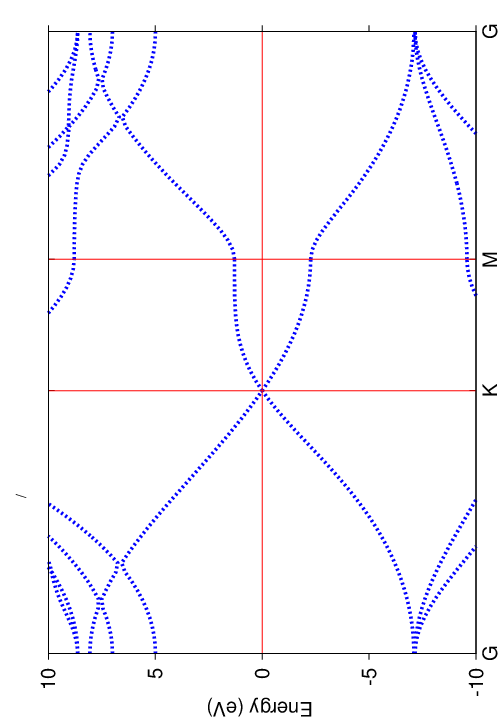}
\includegraphics[width=3.5cm,height=4.5cm,angle=270]{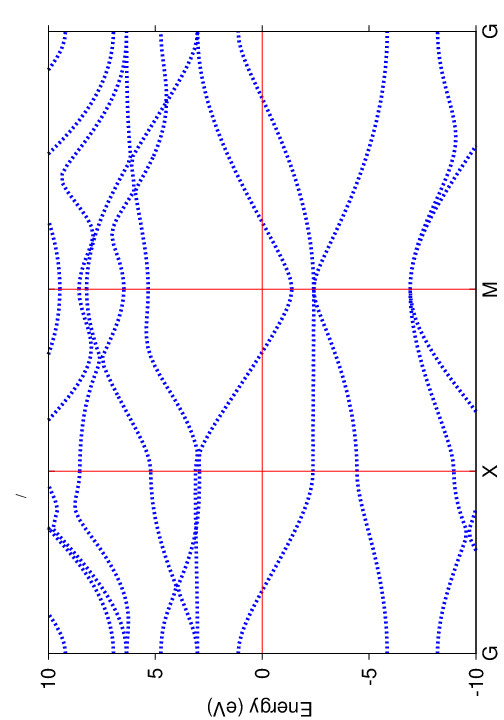}
 \caption{(Color Online) The density of states (DOS) and band structure (BS) of (a) graphene and (b) planar T graphene. The Fermi level is set to 0 eV. The linear dispersion appears due to the crossing of two opposite cones at K point in the band structure (BS) in graphene whereas planar T graphene is metallic. 
\label{Gbs}}
\end{figure}

	Attempts at changing the electronic and magnetic properties of graphene lead to difficulties in maintaining materials stability. Random vacancies  lead to extrinsic type of disorder that we have used here. To understand the effects of random vacancies  in graphene and planar T graphene systems, we first perform band structure and density of states calculations in the ordered systems. First principles calculations were performed on DFT-based tight binding linear muffin tin orbitals (TB-LMTO) within local spin density approximation (LSDA) using von Barth and Hedin (vBH) exchange-correlation functional \cite{vbh}. The $\vec{q}$-space integration was carried out with the Monkhorst-Pack method for the generation of special $\vec{q}$ points in the Brillouin zone. We started our calculation by the relaxation of graphene structure, by minimizing the total energy with respect to the lattice constant. The relaxed  lattice constant came to 2.46 ${\rm\AA}$ and the  C-C bond length  to 1.42${\rm \AA}$. These are in good agreement with ref \cite{para}. To avoid the interactions between two adjacent monolayer graphenes, a vacuum space in Z axis of 15${\rm\AA}$ was chosen  in our numerical simulations. The calculated electronic band structure (BS) along high symmetric $\Gamma-K-M-\Gamma$ direction and density of states (DOS) for pure graphene are as shown in left panel of Fig.\ref{Gbs}. The density of states combines conduction band and valence band forming Dirac point at Fermi level, which are mainly due to the $\pi$ and $\pi$* bonding of $p_z$ states. For such a $sp^2$  bonded structure, the  $\sigma$ bond is responsible for the formation of the underlying framework. The remaining $p_z$ orbital which is roughly vertical to the tetra or hexa-rings, binds covalently with each other and forms $\pi$ and $\pi$* bands. These $\pi$ and $\pi$* are believed to induce the linear dispersion relation at the K point near the Fermi surface leading to its `semi-metallic' behavior. As a result, graphene is a zero-gap semiconductor. In Dirac points the effective mass is equal to zero and the carrier mobility in graphene theoretically diverges. J. H. Chen $\etal$ reported experimentally Carrier mobility in graphene is to be as high as 100000 \cite{mobi}. The results of DFT-based TB-LMTO  band structure (BS) and density of states (DOS) for pure planar T graphene are shown in right panel of Fig \ref{Gbs}. The contribution of $p_z$ states is dominating in total density of states (TDOS) of planar T graphene like that of graphene. The calculated electronic band structure (BS) is along high symmetric $\Gamma-X-M-\Gamma$ direction for both systems. The linear dispersion appears at Dirac point for graphene. Whereas for planar T graphene, the  $\pi$ and $\pi$* does not induce linear dispersion at Fermi level and hence, the band structure becomes metal-like.

\subsection{Electronic structure of graphene and planar T graphene  with random vacancies}

 \begin{figure}[h!]
\centering
{\subfigure{\includegraphics[width=6.5cm,height=6.6cm,angle=270]{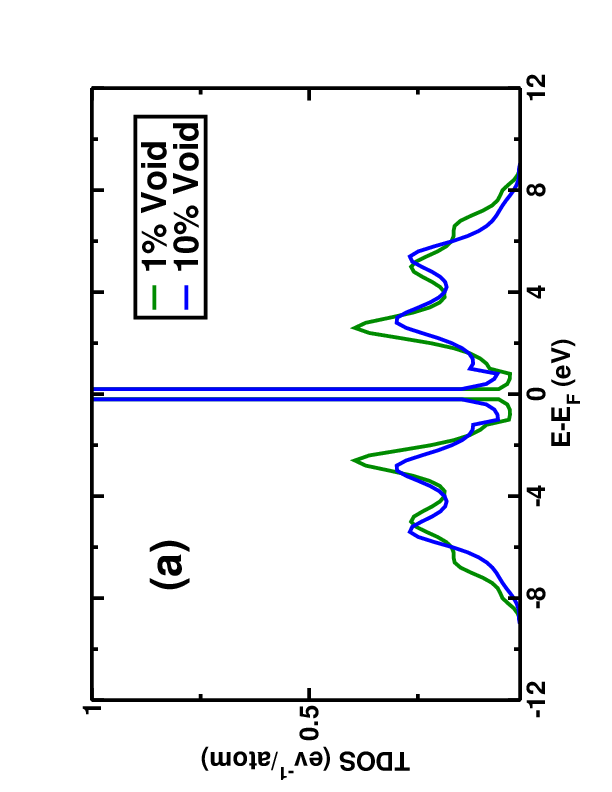}}} \hskip -0.3cm
{\subfigure{\includegraphics[width=6.5cm,height=6.6cm,angle=270]{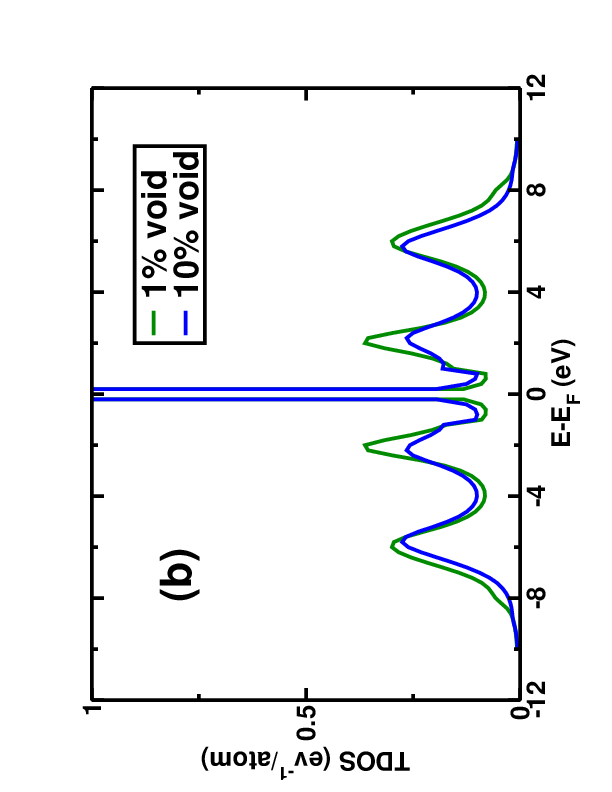}}}
\caption{ (Color Online)  Configuration averaged total density of states (TDOS) for  (a) graphene and (b) T graphene  from real space ASR technique. Dirac peak broadens with increasing disorder strength (vacancies)
\label{tdos}}
\end{figure}

 						We have used augmented real space (ASR) technique to obtain the configuration averaged density of states (ADOS) for disordered systems. We have  1\% and  10\% vacancy concentration both in graphene and planar T graphene. The total density of states (TDOS) of graphene and planar T graphene are depicted in Fig \ref{tdos}.  Disorder induces the formation of the localized states at the Dirac point for both graphene and T graphene. These localized states  get wider with increasing vacancy concentration. The density of states near Dirac point is enhanced due to broadening of the single particle states. If we examine the band edges, we note that whenever there is periodicity at any scale at all, the band edges are sharp and quadratic. Long range disorder leads to band tailing. This disorder induced band tailing has been known for decades and it would be interesting to examine if the tail states are localized or not. In fact, local density of states (LDOS) for the interacting system \cite{PRBq}, \cite{PRBw} can be used in generalized inverse participation ratio (GIPR) to test the signature of localization of the states. It is well known that the single particle wave functions \cite{Jana1} are normally used for non-interacting system to compute this GIPR . Again periodicity at whatever scale leads to structure in the DOS. This is smoothed out by the disorder scattering that induces complex `self-energy'. Occasionally practitioners of super-cell techniques introduce an artificial imaginary part to the energy. This smoothes these structures too, but the procedure is entirely ad hoc. ASR leads to an energy dependent self-energy ($\Sigma(\vec{q})$) systematically. This same $\Sigma(\vec{q})$ broadens the infra-red divergences at the Dirac point. 
 
\subsection{ Effect of random vacancies on the spectral function of graphene and planar T graphene}

\begin{figure}[h!]

\centering
\includegraphics[width=4.0cm,height=4.3cm,angle=270]{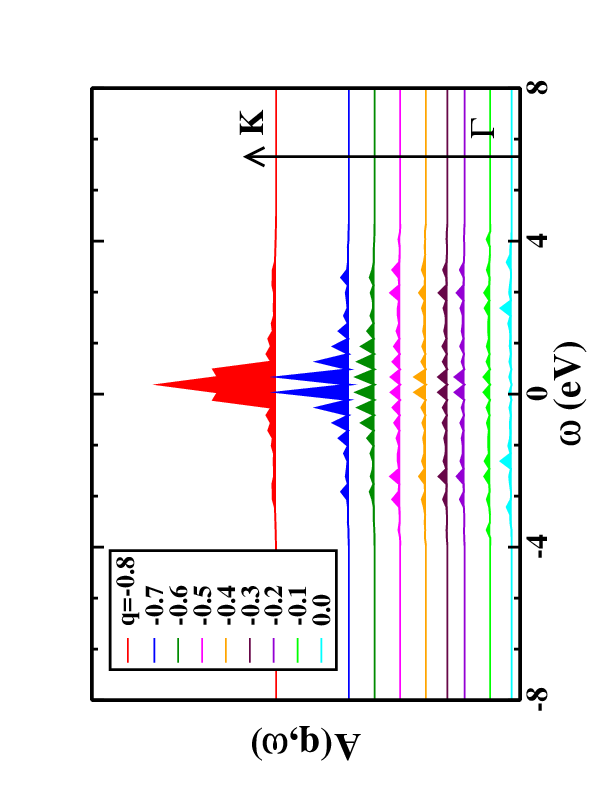}
\includegraphics[width=4.0cm,height=4.3cm,angle=270]{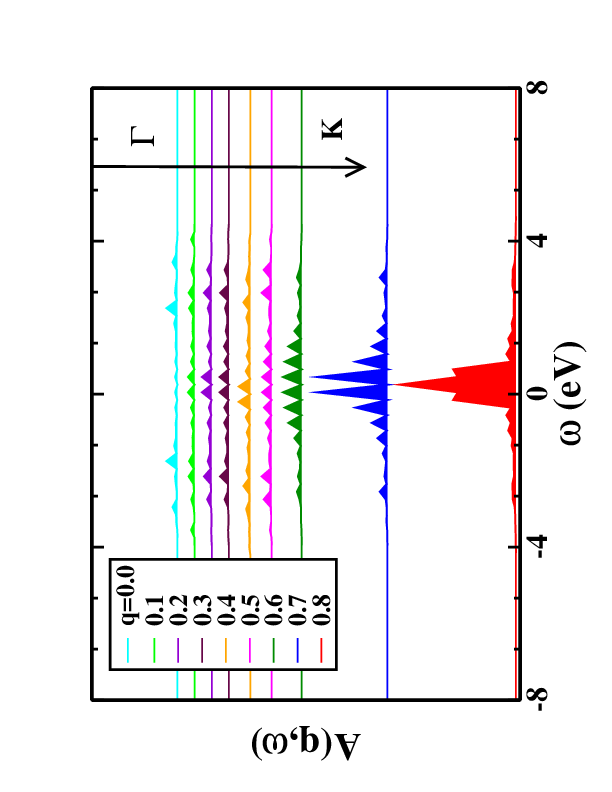}
\caption{ (Color Online) Configuration averaged spectral function of graphene with 10\% vacancy. Wave vector $\vec{q}$ ranges from -0.8 to 0.8 in the first brillouin zone along the symmetric $\Gamma-K$ direction from ASR technique in reciprocal space.
\label{spec1}}
\end{figure}

\begin{figure}[t!]
\centering
{\subfigure{\includegraphics[width=4.0cm,height=4.3cm,angle=270]{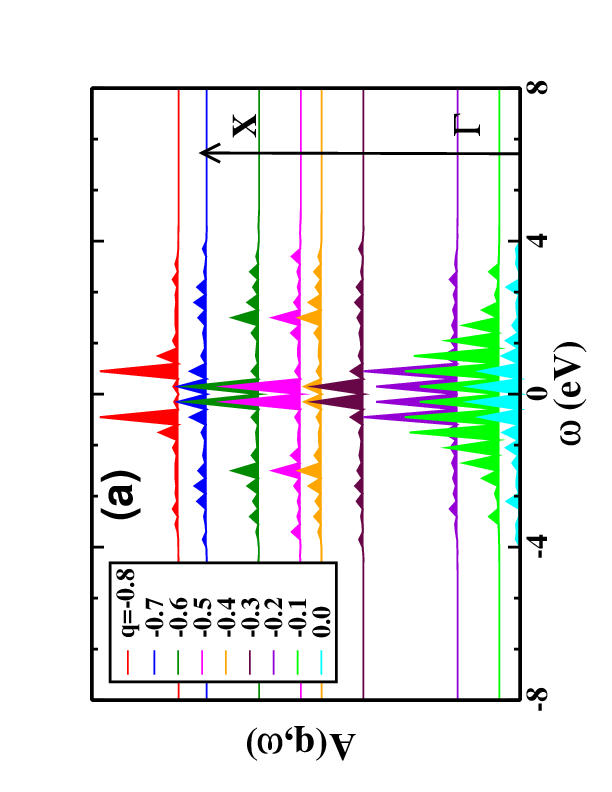}}}
{\subfigure{\includegraphics[width=4.0cm,height=4.3cm,angle=270]{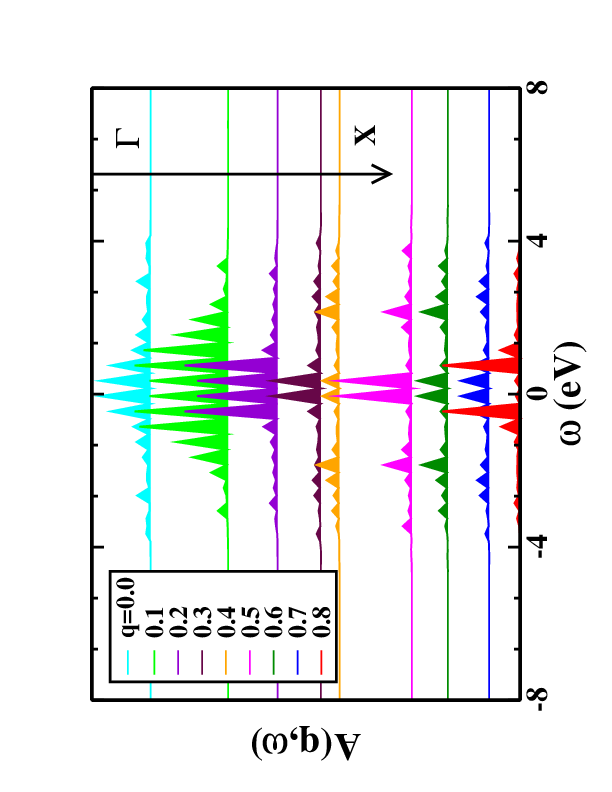}}}\\ 
{\subfigure{\includegraphics[width=4.0cm,height=4.3cm,angle=270]{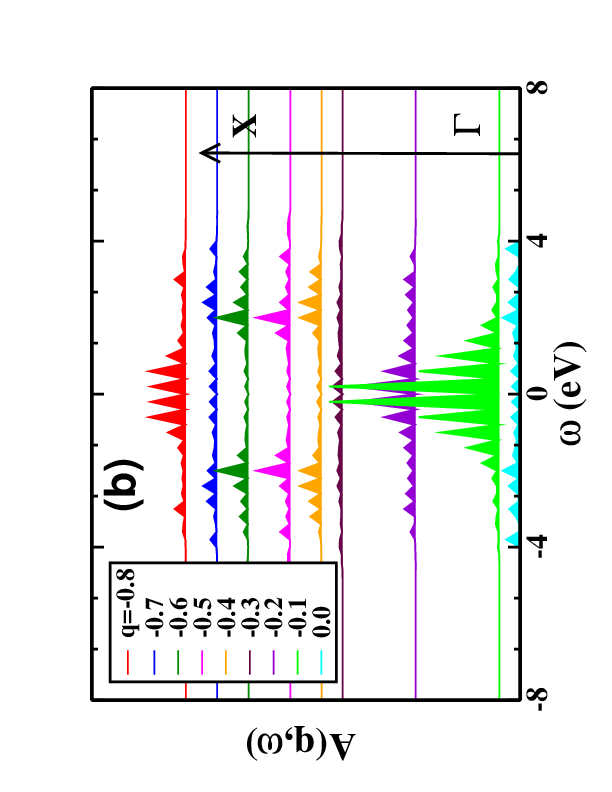}}} 
{\subfigure{\includegraphics[width=4.0cm,height=4.3cm,angle=270]{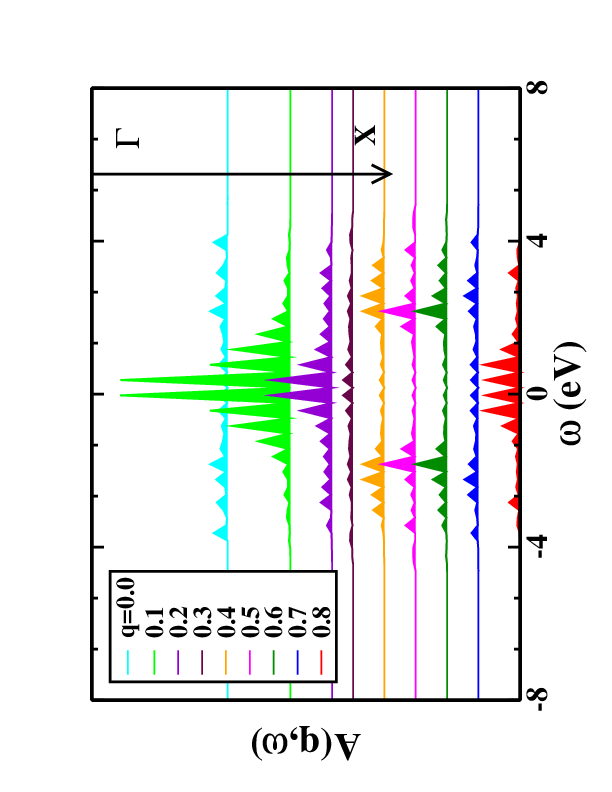}}}\\  
{\subfigure{\includegraphics[width=4.0cm,height=4.3cm,angle=270]{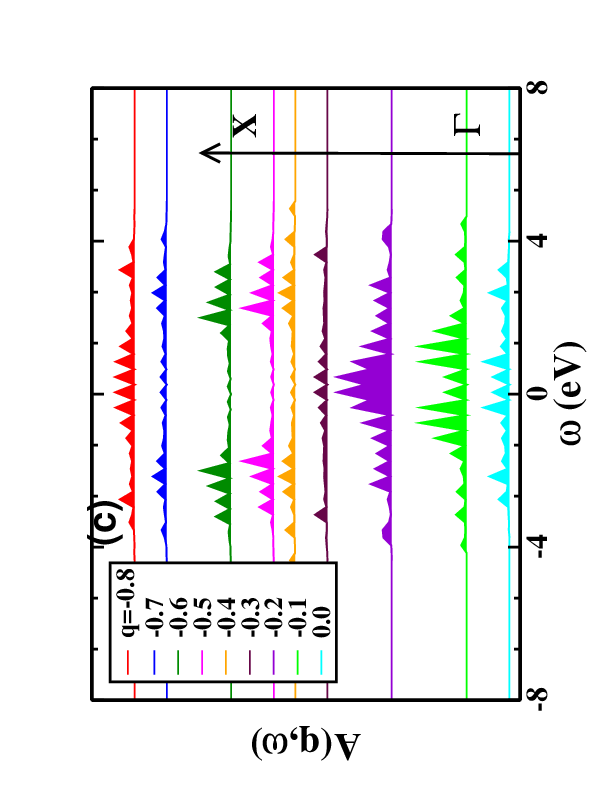}}} 
{\subfigure{\includegraphics[width=4.0cm,height=4.3cm,angle=270]{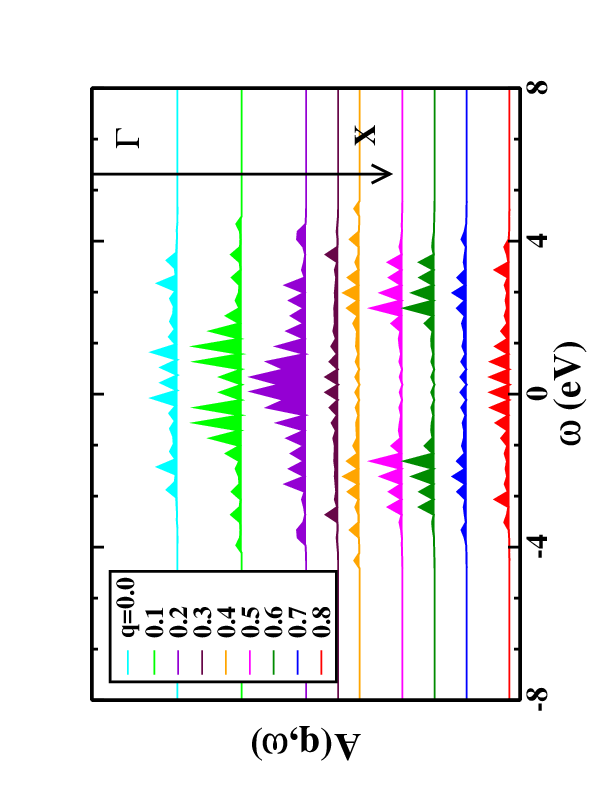}}}\\  
{\subfigure{\includegraphics[width=4.0cm,height=4.3cm,angle=270]{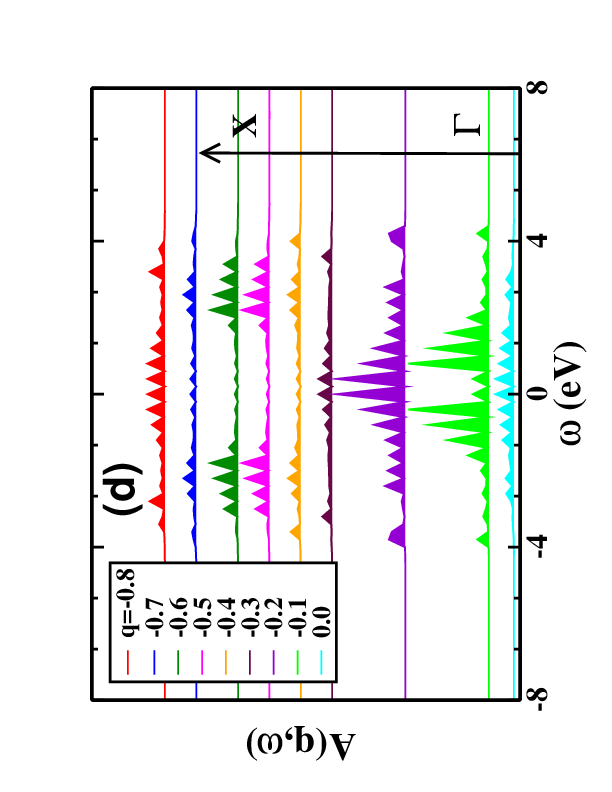}}}
{\subfigure{\includegraphics[width=4.0cm,height=4.3cm,angle=270]{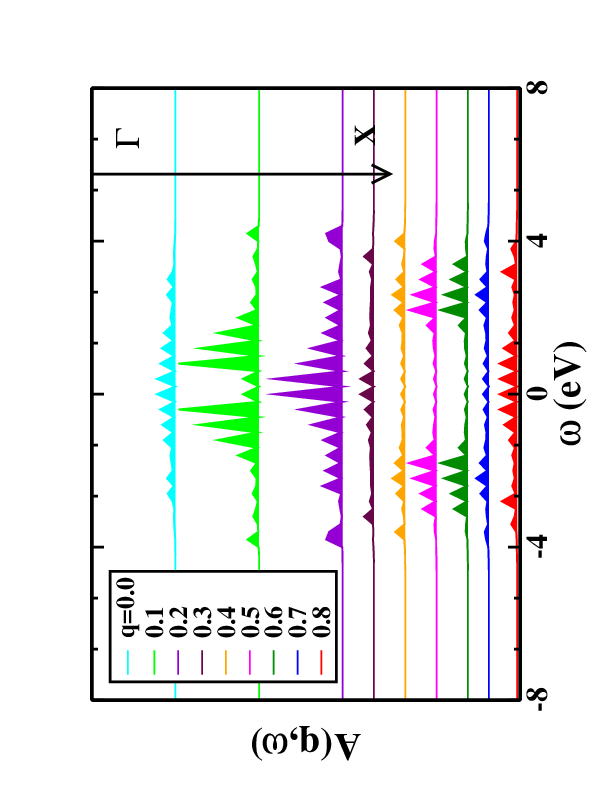}}}
\caption{ (Color Online) Configuration averaged spectral function for T graphene with (a) 1\%, (b) 5\%, (c) 10\%, (d) 11\%  vacancies. Here  the wave vector $\vec{q}$ is varying from -0.8 to 0.8 in the first brillouin zone along the symmetric $\Gamma-X$ direction. The spectral intensity is strongly peaked at $\Lambda$(0.0,0.21) point between $\Gamma-X$ for 10\% vacancy. 
\label{spec2}}
\end{figure}

				We have calculated the configuration averaged spectral function A($\vec{q},\omega$) for a selection of different wave vectors $\vec{q}$ varying from 0.0 to 0.8. Spectral functions are  along the symmetric $\Gamma-K$ direction for graphene with 10 \% vacancy concentation and $\Gamma-X$ direction for that of T graphene  with 1\% to 11\%  vacancy concentrations. Disorder has a large impact on electronic structure including asymmetries in spectral properties as well as in the band structures. In absence of disorder, the spectral function is a bunch of Dirac delta functions. It widens to a Lorenzian with increasing disorder strength. For graphene with random vacancies, the spectral intensity of A($\vec{q},\omega$) is low at $\vec{q}$ = 0.0\AA$^{-1}$ i.e. at $\Gamma$ point, and increases with increasing $\vec{q}$. This is made clear by observing the peaks located at $\vec{q}$ = 0.0 {\AA}$^{-1}$ and 0.6 {\AA}$^{-1}$ as shown in Fig \ref{spec1}. The spectral intensity A($\vec{q},\omega$) is highest at the Dirac point i.e. at $K$(0.0,0.8) point. Whereas for T graphene with 10\% vacancies, spectral intensity is highest at $\vec{q}$ = 0.2 {\AA}$^{-1}$ i.e. at  $\Lambda$(0.0,0.21) point between $\Gamma-X$. If we slowly increase vacancy concentration, at 11\% the spectral intensity of $\Lambda$(0.0,0.21) point decreases as shown in Fig \ref{spec2}. These peak  positions are shifted to lower energies and  exhibit noticeable broadening with increasing wave vector $\vec{q}$ captured by our method. If we increase the disorder concentration the spectral function  A($\vec{q},\omega$) is found to have double peak structure due to scattering induced resonance modes \cite{phonon}, which is reported previously by our group \cite{gdos}.

\subsection{ Effect of random vacancies on the dispersion curves of graphene and planar T graphene : }

\begin{figure*}[h!]
\centering
\subfigure{\includegraphics[width=5.0cm,height=6.0cm,angle=270]{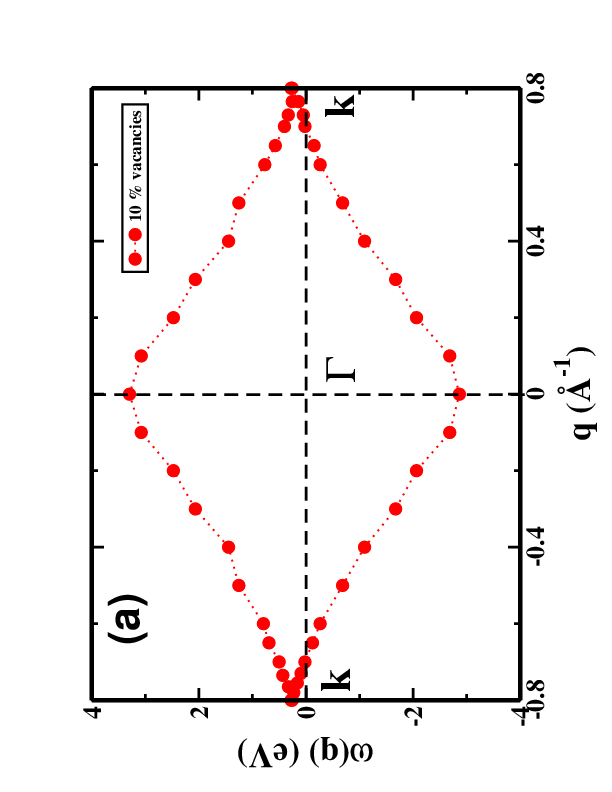}}\hskip -0.2cm
\subfigure{\includegraphics[width=5.0cm,height=6.0cm,angle=270]{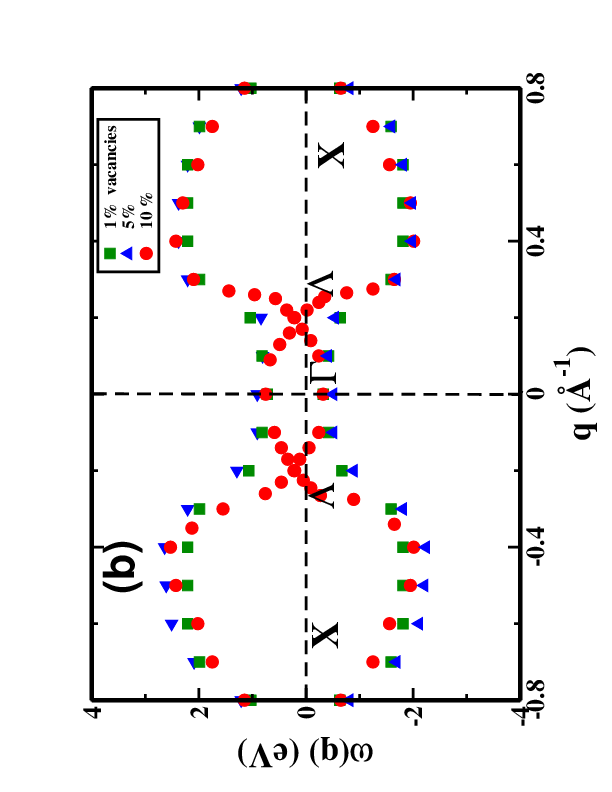}}\hskip -0.4cm
\subfigure{\includegraphics[width=5.0cm,height=6.0cm,angle=270]{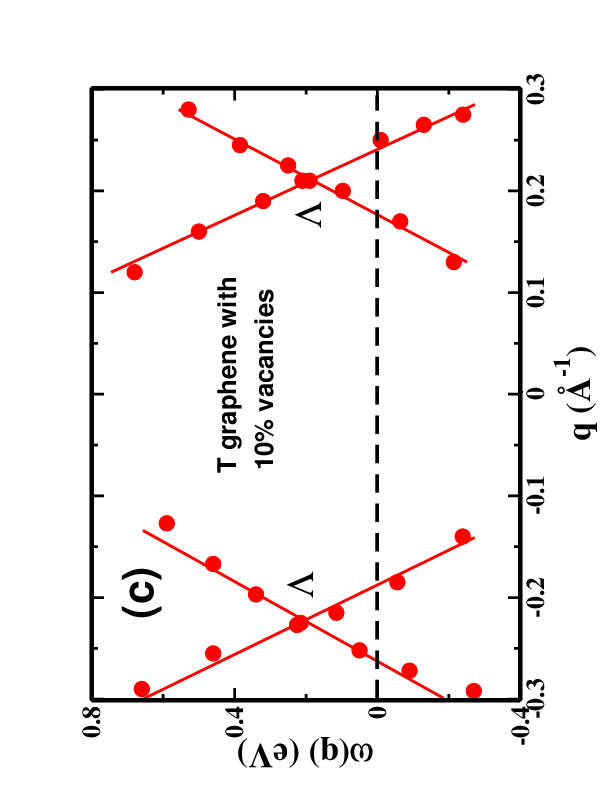}}
\caption{(Color Online) $\pi$ - $\pi*$ band structure (BS) near the Fermi level in the first brillouin zone for both (a) graphene and (b) and (c) planar T graphene with 10\% vacancy. It is obtained by following the peak-finding process described \cite{pk4} in the text. With 10\% vacancy concentration, T graphene induces dirac dispersion like graphene. The Fermi energy $\bf{E_F}$ is at 0 eV. 
\label{bd1}}
\end{figure*}

\begin{figure*}[t!]
\centering
\subfigure{\includegraphics[width=6.5cm,height=6.5cm]{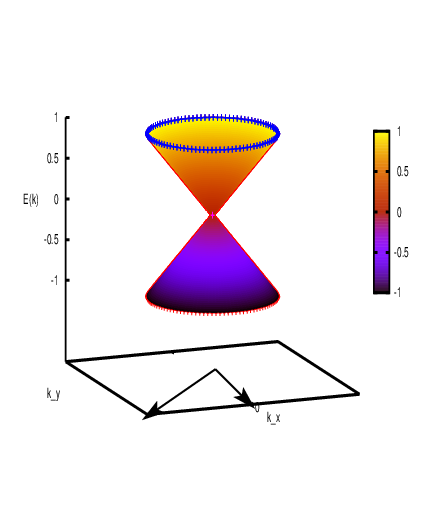}}\hskip 1.0 cm
\subfigure{\includegraphics[width=10.0cm,height=6.0cm]{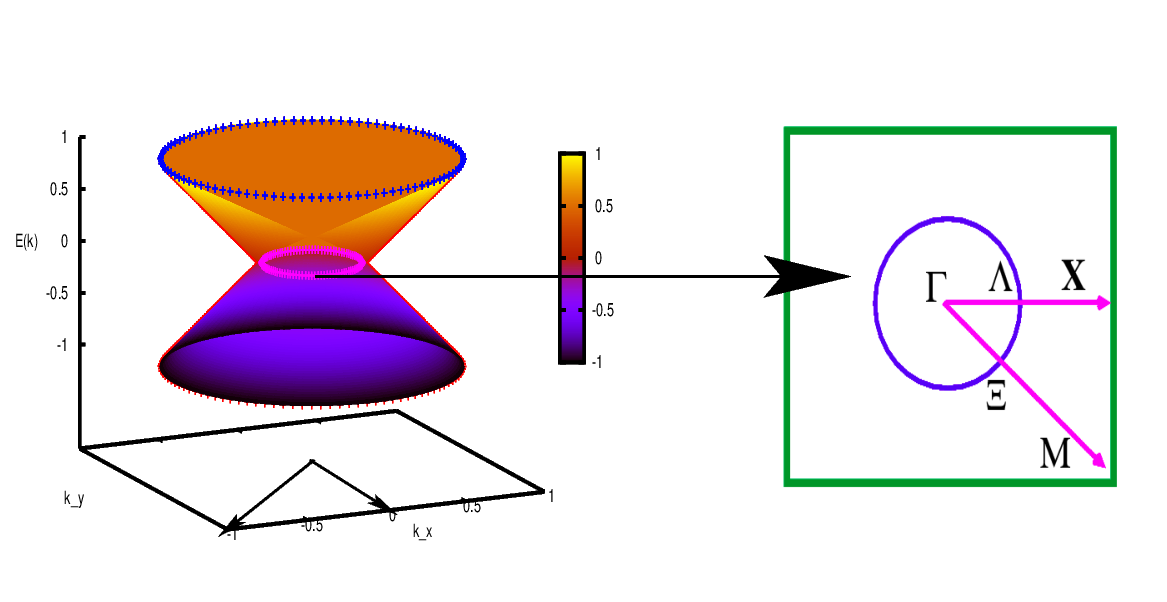}}
\caption{(Color Online) The band dispersion for both graphene (left) and planar T graphene (right) with 10\% vacancy. Graphene has linear dispersion at k points and for T graphene, it is spread out like a loop centered at $\Gamma$. The loop (blue line) is formed by the crossing points : $\Lambda$ and $\Xi$ in the first Brillouin zone (green line). The Fermi energy $\bf{E_F}$ is not at 0 eV. 
\label{bd2}}
\end{figure*}

						We have obtained the band dispersions by identifying the
peak positions of the spectral function A($\vec{q},\omega$) along high symmetric directions in reciprocal space \cite{pk1} - \cite{pk4} for  graphene and T graphene with random vacancies. The calculated dispersion of the $\pi$ - $\pi*$ bands for both  graphene and T graphene with random vacancies, is displayed in Fig \ref{bd1}. For T graphene, Dirac-like band dispersion is obtained with 10\% vacancy concentration. However, we focus on 10\% vacancies for both structures. Both systems show  linear dispersion above the Fermi level. Such asymmetries can be believed as a tendency of more scattering to occur near the resonance energies. The $p_z$ branches of graphene show the typical cusp like behavior at the K point leading to its `semi-metallic' behavior, whereas the Dirac cusp in T graphene is spread out around $\Gamma$ point in reciprocal space. This is also reflected in the TDOS of the two materials. For both disordered graphene and T graphene, the Fermi level is away from the charge neutral Dirac point (taken to be the energy zero) depicted clearly in Fig.\ref{bd1}. Graphene shows linear dispersion at K points near the Fermi level whereas, disordered T graphene shows the linear dispersion around $\Gamma$ point near the Fermi level and forms a loop with tetra-symmetry in Brillouin zone. The two cross points located
at asymmetric positions, $\Lambda$(0.0,0.21) between X and $\Gamma$ and $\Xi$(0.145,0.145) between $\Gamma$ and M in reciprocal space as shown in Fig \ref{bd2}. Yu Liu $\etal$ have reported previously the $\Lambda$(0.0,0.249) and $\Xi$(0.17,0.17) points for buckled T graphene in their communication \cite{hexa2}. The linear dispersion relation near the Fermi level indicates that carriers in  T graphene with random vacancies should be massless and have a Dirac fermionic character like graphene. For such a $sp^2$  bonded structure, it could be graphene or T graphene, the  $\sigma$ bond is responsible for the formation of the underlying framework. The remaining $p_z$ orbital, which is roughly vertical to the tetra or hexa-rings, binds covalently with each other and forms $\pi$ and $\pi$* bands. These $\pi$ and $\pi$* are believed to induce the linear dispersion relation near the Fermi surface. So the emergence of Dirac-like fermions in T graphene is due to $\pi$ and $\pi$* bands, which is similar to that of graphene. Experimentally it is observed that low-buckled Silicene has a similar band structure and Dirac-like fermions with regular hexagonal symmetry \cite{sili1} - \cite{sili2}. Our recent study shows that Dirac fermions can persist in the T graphene without a regular hexagonal symmetry. Experimental investigations predict the dynamic stability and less distortions free formation of T graphene comparing over graphene \cite{distor}. Kotakoski and co-workers \cite{Kota}, Lahiri and co-workers \cite{Lahi} have experimentally obtained such a 1D carbon structure consisting alternately of carbon tetrarings (C4) and octarings (C8). Physically, it is well khown that  symmetry breaking or small alteration in structure leads to drastic changes in properties in many compounds. We anticipate that introducing random vacancies,
as described above, leads to rippling in planar T graphene. This changes the behavior of planar T graphene and linear dispersion appears to be similar to that of buckled T graphene.

\subsection{ Disorder induced relaxation times of graphene and planar T graphene : }

\begin{figure}[h!]
\centering
\subfigure{\includegraphics[width=7.5cm,height=8.0cm,angle=270]{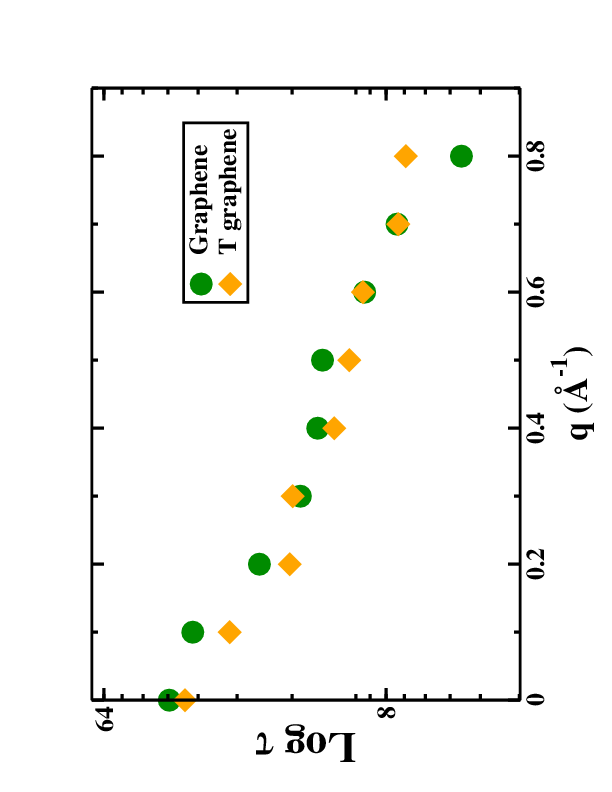}}
\caption{(Color Online) Single particle relaxation time with 10\% vacancy for graphene and T graphene in the first brillouin Zone. Relaxing modes are labelled by different wave vector q (${\AA}^{-1}$). 
\label{relax}}
\end{figure}

						Disorder (vacancy) in graphene and T graphene leads to two distinct momentum relaxation times ; the transport relaxation time and the quantum lifetime or the single-particle relaxation time. Our recent study does not take into account the transport relaxation time, we only calculate the single-particle relaxation time $\tau(\vec{q})$. From a many body theory, the single particle relaxation time $\tau(\vec{q})$  is calculated from the electron self energy of the coupled electron-impurity system. We obtain disorder induced lifetime $\tau(\vec{q})$ of a $ \vec{q} $ labelled quantum state from the fourier transform of the configuration averaged self energy function. Finally, in Fig.\ref{relax}  we show our calculated single particle relaxation time $\tau(\vec{q})$  as a function of $\vec{q}$ (${\AA}^{-1}$) in the first  brillouin zone for both  graphene and T graphene. The relaxing modes or patterns are labelled by $\vec{q}$, such that the average `size' of the mode is $~O(q^{-1})$. We define the lifetime $\tau(\vec{q})$ as the time in which the disordered induced self energy amplitude drops to its $1/e$ value. We note that  Log $\tau(\vec{q})$ decreases almost linearly with increasing the wave vector $\vec{q}$. Previously it is reported that for short ranged scatterers with which we are dealing here, the ratio of the scattering time to the single particle relaxation time is always smaller than 2 \cite{relax}. In experiment, the ratio of the transport time to the single particle time has been found to be smaller than 1 in Si-MOSFET-systems. Thus, in Si-MOSFET-system, the short-ranged scattering (such as interface roughness scattering) dominates. To date, most experimental studies have focused on the transport scattering time. Both the single particle relaxation time and transport scattering time are important for carrier mobilities ($ \mu $ ) in 2D graphene like systems. This ratio of two types of scattering time is needed for describing scattering mechanism in such 2D systems. It is important to note that the single particle relaxation time $\tau(\vec{q})$ is accessible to neutron scattering experiments.

\section{Conclusion}
 
 In conclusion, we have systematically studied the effect of disorder (random vacancies) on pristine graphene and planar T graphene from both real and reciprocal space approaches. We have also proposed a new approach  using augmented space (ASR) technique for handling disorder without any external parameter fitting in the the complex part of energy. This will help to get disorder induced lifetimes for the study of other disorder related effects in 2D materials. The single particle relaxation time  $\tau(\vec{q})$ has been obtained from the broadening of quantum states due to disorder. This study will guide us to understand the nature of disordered scattering in graphene and T graphene. We also examine the energy band dispersion  and the density of states for these two subtances. The density of states get enhanced near the Dirac point due to broadening of the single particle states for both structures. Graphene like linear dispersion is appeared in planar T graphene in the vacancy concentration range very close to 10\%. We should note that no honeycomb structure is needed to produce Dirac point-like dispersion. In nano-technological application, the synthesis of two dimensional structures containing carbon tetrarings (C4) and octarings (C8) by defect engineering play an important role. Up to now the effects of disorder in planar T graphene has not been well studied. We conclude by emphasizing that  our theoretical study will provide a reference and detailed useful insight for building interesting nano materials \cite{vv} - \cite{zhou} based on 2D carbon allotropes without hexagonal symmetry.

\section*{Acknowledgements}
This work was done under the Hydra Collaboration between our institutes. Authors are deeply grateful to M. Kabir (IISER Pune, India) for a fruitful discussion and Dhani Nafday, Tanusri Saha-Dasgupta for providing us with the Hamiltonian parameters calculated by NMTO. We would like to thank Prof. O.K. Anderson, Max Plank Institute, Stuttgart, Germany, for his kind permission to use TB-LMTO code developed by his group.


\begin{thebibliography}{99}
\bibitem{2d} Novoselov {\etal} \textit{Science} \textbf {306} 666 (2004)
\bibitem{massless} K Novoselov, A Geim, S Morozov, D Jiang, A Firsov \textit{Nature (London)} \textbf {438} 197 (2005) 
\bibitem{apl1} A Neto, F Guinea, N Peres, K Novoselov, A Geim \textit{ Rev. Mod. Phys} \textbf {81} 109 (2009)
\bibitem{apl2} E Kan, Z Li, J Yang \textit{NANO} \textbf {3} 433 (2008)
\bibitem{apl3} P Avouris, Z Chen, V Perebeinos \textit{Nature Nanotech} \textbf {2} 605 (2007)
\bibitem{apl4} M Dragoman, D Dragoman \textit{Prog. Quantum Electronics} \textbf {33} 165 (2009)
\bibitem{apl5} A Castro Neto \textit{Mater. Today} \textbf {13} 1 (2010)
\bibitem{hexa} A Enyashin and A Ivanovskii \textit{Phys. Status Solidi B}\textbf {248} 1879 (2011)
\bibitem{hexa1}S Choi, S Jhi, and Y Son \textit{Phys. Rev. B} \textbf {81} 081407 (2010)
\bibitem{hexa2}{\color{blue}{Yu Liu {\etal} \textit{Phys. Rev. Lett} \textbf {108} 225505 (2012)}}
\bibitem{kim} B Kim, J Jo, and H Sim \textit{Phys. Rev. Lett} {\bf{110}} 029601 (2013)
\bibitem{huang} H Huang, Y Li, Z Liu, J Wu, and W Duan \textit{Phys. Rev. Lett}
{\bf{110}} 029603 (2013)
\bibitem{chun} Chun-Sheng Liu, Ran Jia, Xiao-Juan Ye, and Zhi Zeng \textit{The journal of chemical phys} {\bf{139}} 034704 (2013)
\bibitem{pala} J. J. Palacios, J. Fern\"andez-Rossier, and L. Brey \textit{Phys. Rev. B} {\bf 77} 195428 (2008) 
\bibitem{oleg1} Oleg V. Yazyev and Lothar Helm \textit{Phys. Rev. B} {\bf 75} 125408 (2007) 
\bibitem{oleg2} Oleg V. Yazyev \textit{Phys. Rev. Lett.} {\bf 101} 037203 (2008)    
\bibitem{am2} A Mookerjee   \textit{J. Phys. C : Solid State Phys} {\bf 8}   29 (1975)
\bibitem{am3} A Mookerjee   \textit{J. Phys. C : Solid State Phys} {\bf 8} 1524 (1975)
\bibitem{am4} A Mookerjee   \textit{J. Phys. C : Solid State Phys} {\bf 8} 2688 (1975)
\bibitem{am5} A Mookerjee   \textit{J. Phys. C : Solid State Phys} {\bf 8} 2943 (1975)
\bibitem{am1} A Mookerjee   \textit{J. Phys. C : Solid State Phys} {\bf 6} 1340 (1973)
\bibitem{config} S Chowdhury {\etal} \textit{ Indian J Phys} DOI : 10.1007/s12648-015-0789-2  arXiv:1404.6778v2 (2015) 
\bibitem{nlcpa} D Biava  {\etal} \textit{Phys. Rev. B} {\bf 72} 113105 (2005)
\bibitem{sqs} A Zunger  {\etal} \textit{Phys. Rev. Lett} \textbf{65} 353 (1990)
\bibitem{lsms} Y Wang {\etal} \textit{Phys. Rev. Lett} \textbf{75} 2867 (1995)
\bibitem{tca} R Mills and P Ratanavararaksa \textit{Phys. Rev. B} \textbf{18} 5291 (1978)
\bibitem{icpa} S Ghosh, P Leath, and M Cohen \textit{Phys. Rev. B} \textbf{66} 214206
(2002)
\bibitem{peres} N Peres, F Guinea, and A Castro Neto \textit{Phys. Rev. B} \textbf {73}
125411 (2006)
\bibitem{inv} S Wu  {\etal} \textit{Phys. Rev. B} \textbf {77} 195411 (2008)
\bibitem{Doniach} S Doniach, E Sondheimer \textit{Green’s Functions for Solid State Physicists Imperial College Press} (1998) \textbf {chapter-V,137}
\bibitem{Doniach1} S Doniach, E Sondheimer \textit{Green's Functions for Solid State Physicists} Imperial College Press (1998) chapter-V,114-140
\bibitem{sw1} Suman Chowdhury {\etal} \textit{Physica E: Low-dimensional Systems and Nanostructures} \textbf {61} 191 (2014) DOI : http://dx.doi.org/10.1016/j.physe.2014.04.002
\bibitem{sw2} Ransell D'Souza, Sugata Mukherjee, Tanusri Saha-Dasgupta \textit{Journal of Alloys and Compounds} \textbf {708} 437 (2017) DOI :  http://dx.doi.org/10.1016/j.jallcom.2017.03.006
\bibitem{cube}  Tanusri Saha and Abhijit Mookerjee \textit{J. Phys. C : Solid State Phys} \textbf {8} 2915 (1995)
\bibitem{exp1} P Esquinazi {\etal} \textit{Phys. Rev. B} \textbf{66} 024429 (2002)
\bibitem{exp2} P Esquinazi  {\etal} \textit{Phys. Rev. Lett} \textbf{91} 227201 (2003)
\bibitem{exp3} A Rode  {\etal} \textit{Phys. Rev. B} \textbf{70} 054407 (2004)
 \bibitem{exp4} T Makarova and F Palacio, eds., Carbon-Based Magnetism: an
overview of met l free carbon-based compounds and materials
(Elsevier, Amsterdam, 2005)
\bibitem{exp5} T Matsui  {\etal} \textit{Phys. Rev. Lett} \textbf{94} 226403 (2005)

\bibitem{recur} R Haydock , V Heine  and  M Kelly \textit{J. Phys. C : Solid State Phys} \textbf{5} 2485 (1972)
\bibitem{recur1} Haydock, R in  ``Solid State Physics", edited by H Ehrenreich,F Sietz, D Turnbull, Academic, New York, 35 (1980)
\bibitem{haydock} R Haydock and C M M Nex \textit{Phys. Rev. B} \textbf{74} 20521 (2006)
\bibitem{nex} M Luchini and C M M Nex \textit{J. Phys. C : Solid State Phys}. \textbf{20}, 3125 (1987)
\bibitem{pettifor} N Beer, D Pettifor, P Phariseau and W Temmerman \textit{(eds.) Electronic Structure of Complex Systems} New York : Plenum Press
\bibitem{vs1} V Viswanath and G M\"uller \textit{The Recursion Method, Applications to Many-Body Dynamics} Germany : Springer-Verlag (1994)
\bibitem{vs2} V Viswanath, Shu Zhang, Gerhard Müller, and Joachim Stolze \textit{Phys. Rev. B} \textbf{51} 368 (1995)
\bibitem{vs3} V Viswanath and G M\"uller \textit{J.Appl.Phys} \textbf{67} 5486 (1990)
\bibitem{vs4} Hans Beck, Gerhard Müller \textit{Solid State Communications}  \textbf{43(6)} 399-403 (1982)
\bibitem{bct1} W Mao {\etal} \textit{Science} \textbf {302} 425 (2003)
\bibitem{bct2} K Umemoto, R Wentzcovitch, S Saito, and T Miyake \textit{ Phys. Rev. Lett} \textbf {104} 125504 (2010).
\bibitem{ASR} A  Mookerjee, D D Sarma (Eds.), Electronics Structure of Clustures, Surfaces and Disordered Solids Taylors Francis  (2003)
\bibitem{sg} Shreemoyee Ganguly {\sl Thesis} Calcutta University (2014)
\bibitem{sg1} S Ganguly  {\etal} \textit{Phys. Rev. B} \textbf{83} 094407 (2011)
\bibitem{vbh} U von Barth and L Hedin \textit{J. Phys. C} \textbf{5} 1629 (1972)
\bibitem{para} Avouris P, Graphene: Electronic and Photonic Properties and Devices. \textit{Nano Lett}
\bibitem{mobi} J H Chen  {\etal} \textit{Nature Nanotechnology} \textbf{3} 206 (2008)
\bibitem{PRBq} N C Murphy, R Wortis and W A Atkinson \textit{Phys. Rev. B} \textbf{83} 184206 (2011)
\bibitem{PRBw} E Cuevas \textit{Phys. Rev. B} \textbf{66} 233103 (2002)
\bibitem{Jana1} D Jana \textit{Ind. Jour. Phys} \textbf{74A} 547 (2000)
\bibitem{phonon} Y V Skrypnyk and V M Loktev, \textit{Phys. Rev. B} \textbf {75} 245401 (2007)
\bibitem{gdos} {\color{blue}{Aftab Alam , Biplab Sanyal , Abhijit Mookerjee \textit{ Physical Rev. B } \textbf {86} 085454 (2012)}}
\bibitem{pk1} B Skubic, J Hellsvik, L Nordström, and O Eriksson  \textit{J. Phys.Condens.Matter} \textbf{20} 31 (2008)
\bibitem{pk2} X Tao, D Landau, T Schulthess, and G Stocks, \textit{Phys.Rev. Lett} \textbf{95} 087207 (2005)
\bibitem{pk3} K Chen and D Landau \textit{Phys. Rev. B}  \textbf{49} 3266 (1994)
\bibitem{pk4}{\color{blue}{ Anders Bergman  {\etal} \textit{Phys. Rev. B} \textbf{81} 144416 (2010) DOI: 10.1103/PhysRevB.81.144416}}
\bibitem{sili1} S Cahangirov  {\etal} \textit{Phys. Rev. Lett} \textbf{102} 236804 (2009)  
\bibitem{sili2} P De Padova \textit{Appl. Phys. Lett} \textbf{96} 261905 (2010)
\bibitem{distor} J C Meyer  {\etal}\textit{ Nature (London)}
\textbf{446} 60 (2007)
\bibitem{Kota} J Kotakoski  {\etal} \textit{Phys. Rev. Lett} \textbf{106} 105505 (2011)
\bibitem{Lahi} J Lahiri, Y Lin, P Bozkurt, I I Oleynik and M. Batzill \textit{Nature Nanotech} \textbf{5} 326 (2010)
\bibitem{relax} E H Hwang, S Das Sarma \textit{Phys. Rev. B} \textbf{77} 195412 (2008)
\bibitem{vv} V V Cheianov, V Fal’ko, and B L Altshuler \textit{Science} \textbf{315} 1252 (2007)
\bibitem{zhou} M Zhou, Y Lu, Y Cai, C Zhang, and Y P Feng \textit{Nanotechnology} \textbf{22}
385502 (2011)

 
\end{thebibliography}
\end{document}